\documentclass[11pt,a4paper]{article}
\usepackage[utf8]{inputenc}
\usepackage[T1]{fontenc}
\usepackage[english]{babel}
\usepackage{here}
\usepackage{graphicx}
\usepackage{subcaption}
\usepackage{eurosym}
\usepackage{amsmath}
\usepackage{mathtools}
\usepackage{braket}
\usepackage{bm}
\usepackage{amssymb}
\usepackage{latexsym}
\usepackage{xcolor}
\usepackage{mathrsfs}
\usepackage{caption}
\usepackage{float}
\usepackage[compact]{titlesec}

\usepackage{hyperref}
\usepackage[style=numeric,sorting=none,url=true,doi=true,backend=bibtex]{biblatex}
\usepackage{cleveref}
\usepackage[switch]{lineno}
\usepackage[font=footnotesize,labelfont=bf]{caption}
\usepackage[hmargin=2cm,vmargin=2cm,rmargin=2cm, lmargin=2cm]{geometry}
\hypersetup{colorlinks=true,citecolor=red,filecolor=blue,linkcolor=blue,urlcolor=blue}
\DeclareUnicodeCharacter{FB01}{fi}
\DeclareUnicodeCharacter{200B}{}

\def\keywordname{{\bfseries \emph{Keywords}}}%
\def\keywords#1{\par\addvspace\medskipamount{\rightskip=0pt plus1cm
\def\and{\ifhmode\unskip\nobreak\fi\ $\cdot$
}\noindent\keywordname\enspace\ignorespaces#1\par}}
\begin{document}
\renewcommand{\figurename}{\textbf{Fig.}}
\renewcommand{\tablename}{\textbf{Table}}
\newcommand{\un}{\mathds{1}}
\newcommand{\1}{\mathbbm{1}}
\newcommand{\vect}[1]{\boldsymbol{#1}}
\newcommand{\era}{\end{array}}
\newcommand{\beq}{\begin{equation}}
\newcommand{\eeq}{\end{equation}}
\newcommand{\beqar}{\begin{eqnarray}}
\newcommand{\eeqar}{\end{eqnarray}}
\newcommand{\lb}{\label}
\thispagestyle{empty}
\baselineskip=18pt
\medskip
\begin{center}
~~~~~~~~~~~~~~~
\\
\vspace{2cm}
\noindent { {\textbf{ Dimeric Perylene-Bisimide Organic Molecules: Fractional-Time Control of Quantum Resources }}}\\

\vspace{0.7cm}
\noindent
\vspace{0.5cm}
{\small Abdessamie Chhieb $^{a,c}$}{\footnote{E-mail:
\textsf{\href{mailto:chhiebabdessamie@gmail.com }{chhiebabdessamie@gmail.com  }}}}, {\small Chaimae Banouni $^{a,c}$}{\footnote{E-mail:
\textsf{\href{mailto:banounichaymae1@gmail.com }{banounichaymae1@gmail.com  }}}}, {\small Sliha Abdessamie $^{a,c}$}{\footnote{E-mail: \textsf{\href{mailto:salihaabdessamia67@gmail.com}{salihaabdessamia67@gmail.com}}}} and 
{\small Mohamed Ouchrif $^{a,c}$}{\footnote{E-mail:
\textsf{\href{mailto:ouchrif@gmail.com }{ouchrif@gmail.com  }}}},

\noindent $^{a}${{\footnotesize   Laboratory of Theoretical Physics, Particles, Modeling and Energies, Faculty of Sciences,\\ Mohamed First University, Oujda,  Morocco}}\\[0.5em]

\noindent $^{c}${{\footnotesize   National Institute For Particle Physics and Applications (NIPPA), Oujda, Morocco}}\\[0.5em]
\end{center}
\vspace{0.5cm}

\begin{abstract}
In this work, we explore the dynamics of quantum correlations, namely coherence, entanglement, and nonlocality associated with a Bell state, in a dimeric arrangement of organic PBI molecules, mediated by dipole-dipole interactions, under time-fractional dynamics. Within the framework of the time-fractional Schrödinger equation (TFSE) with Caputo fractional derivatives, we explore system dynamics for different values of the fractional order $\tau$, transition energies, interaction strength, and purity $p$. We employ the relative entropy of coherence, logarithmic entanglement entropy and concurrence, and CHSH inequality to estimate system dynamics associated with coherence, entanglement, and nonlocality, respectively. These findings highlight the role of the fractional order $\tau$ in system dynamics, including memory effects and relaxation, and thereby bring together ideas from fractional calculus and quantum information theory perspectives and discuss methodologies to control and utilize these molecular quantum correlations.
\end{abstract}

\vspace{0.5cm}
\newpage
\section*{Introduction}
Molecular quantum systems, especially synthetic organic chromophores, have been identified as promising candidates for quantum information processing at ambient temperature. Of these, PBI dimers, which consist of two covalently bonded PBI molecules, are especially noteworthy due to their coherent excitonic properties, which allow for the study of quantum correlations on the single-molecule scale ~\cite{Hildner2011, Gorman2015, Steiner2018}. Recent achievements in quantum control and the implementation of quantum gates on individual organic molecules ~\cite{Brixner2005, Pflumm2015} promise scalable quantum technology on the molecular scale. Quantum technology, including quantum computation, quantum communication, and quantum sensing, is based on non-classical resources such as quantum coherence, quantum entanglement, and quantum Bell nonlocality, which provide a basis for quantum protocols that are classically unfeasible, such as quantum teleportation, quantum dense coding, and quantum key distribution ~\cite{nielsen2010quantum, Horodecki2009}. Yet, the application of quantum technology for scalable applications remains hindered by the interaction of quantum systems with the environment, leading to decoherence and the destruction of quantum correlations on short timescales ~\cite{nielsen2010quantum,preskill2018quantum}.\\

Fractional quantum mechanics provides a powerful theoretical framework for describing complex quantum dynamics characterized by memory effects and long-range temporal correlations. The fractional Schrödinger equation, originally introduced by Laskin~\cite{laskin2002fractional}, and its time-fractional extension (TFSE)~\cite{naber2004time, Tarasov2011}, incorporate fractional derivatives to model non-Markovian and dissipative quantum dynamics~\cite{Hilfer2000}. The TFSE has been successfully applied to a wide range of physical systems, including optical signal processing~\cite{liu2023experimental}, anomalous relaxation phenomena~\cite{aceves2022spatio}, and fractional quantum systems such as the Bohr atom and nonlocal oscillators~\cite{laskin2002fractional}. Unlike the standard Schrödinger equation, which describes a purely unitary and Markovian evolution governed by a first-order time derivative, the TFSE constitutes an effective phenomenological framework for open quantum systems in regimes where conventional Markovian master equations lose their validity~\cite{breuer2002theory}. The use of the Caputo fractional derivative allows for physically meaningful initial conditions~\cite{Hilfer2000, podlubny1998fractional}, while the resulting solutions expressed in terms of Mittag--Leffler functions naturally account for non-exponential, power-law relaxation dynamics~\cite{naber2004time, Tarasov2011, liu2023experimental}. Within this framework, the fractional order parameter $\tau$ acts as a quantitative measure of memory strength and non-Markovianity in the system--environment interaction~\cite{chhieb2024metrological, chhieb2025fractional, chhieb2025time, abdessamie2025non, banouni2025thermal}.\\

In this work, we investigate the dynamics of quantum coherence, entanglement—quantified by the logarithmic negativity~\cite{vidal2002computable, plenio2005logarithmic}—and Bell nonlocality in a PBI dimer described by a time-fractional Schrödinger equation. In our work, we have shown that the fractional parameter $\tau$ is an effective control parameter for controlling the temporal evolution of correlations in order to have a transition from exponential decay to algebraic decay, thus providing new insight for maintaining the quantumness of molecular systems in non-Markovian environments.\\

We divide our paper as follows: Section~\ref{sec3} presents the physical model of PBI dimer, dipole-dipole interaction Hamiltonian, deriving an effective Hamiltonian, as well as deriving a time fractional Schrödinger equation. Section~\ref{sec4} shows our results of analysis and computation regarding evolution of quantum resources varying with fractional parameters. Finally, Section~\ref{sec5} summarizes our findings, discusses their implications for controlling quantum resources in molecular systems, and outlines future research directions at the interface of fractional quantum mechanics and quantum information science.

\section{Quantum resources }
\label{sec2}

\subsection{Bell Nonlocality}

A quantum state of two qubits, denoted as \(\varrho\), can be expressed in the Hilbert-Schmidt formalism as:
\begin{align}
\varrho = \frac{1}{4} \left[ \mathbb{I}_2 \otimes \mathbb{I}_2 + \vec{u} \cdot \vec{\tau} \otimes \mathbb{I}_2 + \mathbb{I}_2 \otimes \vec{v} \cdot \vec{\tau} + \sum_{k,l} n_{kl} \tau_k \otimes \tau_l \right],
\end{align}
where \(\mathbb{I}_2\) represents the \(2 \times 2\) identity matrix, \(\vec{u}\) and \(\vec{v}\) are Bloch vectors describing local properties, and \(n_{kl} = \mathrm{Tr}(\varrho \, \tau_k \otimes \tau_l)\) are the elements of the correlation tensor \(N\).

The phenomenon of quantum nonlocality in two-qubit systems can be examined through the violation of Bell-type inequalities. A prominent example is the \(\mathrm{CHSH}\) inequality. The Bell operator associated with the \(\mathrm{CHSH}\) inequality is given by:
\begin{align}
\mathcal{Q} = X_1 \otimes Y_1 + X_1 \otimes Y_2 + X_2 \otimes Y_1 - X_2 \otimes Y_2,
\end{align}
where:
\begin{align}
X_i = \vec{u}_i \cdot \vec{\tau}_X = u_i^x \tau_X^x + u_i^y \tau_X^y + u_i^z \tau_X^z, \\
Y_j = \vec{v}_j \cdot \vec{\tau}_Y = v_j^x \tau_Y^x + v_j^y \tau_Y^y + v_j^z \tau_Y^z.
\end{align}
Here, \(\vec{u}_i = (u_i^x, u_i^y, u_i^z)\) and \(\vec{v}_j = (v_j^x, v_j^y, v_j^z)\) are unit vectors in real space, and \(\tau_{X/Y}^{x,y,z}\) represent the Pauli matrices.

The \(\mathrm{CHSH}\) inequality can be mathematically expressed as:
\begin{align}
\mathbb{Q} = |\langle \mathcal{Q} \rangle_\varrho| = |\mathrm{Tr}(\varrho \mathcal{Q})| \leq 2.
\end{align}

For a given quantum state \(\varrho\), the maximum expectation value of \(\langle \mathcal{Q} \rangle_\varrho\) over all possible measurement configurations is denoted by \(\langle \mathrm{CHSH} \rangle_\varrho\). It is defined as~\cite{bell1964einstein}:
\begin{align}
\langle \mathrm{CHSH} \rangle_\varrho = \max_{X_i, Y_j} \mathrm{Tr}(\varrho \mathcal{Q}) = \sqrt{\nu_1 + \nu_2},
\end{align}
where \(\nu_1\) and \(\nu_2\) are the two largest eigenvalues of the matrix \(N^\dagger N\). Here, \(N^\dagger\) is the Hermitian conjugate (transpose and complex conjugate) of \(N\), and \(N^\dagger N\) is a \(3 \times 3\) matrix.

\subsection{ Logarithmic negativity}
Logarithmic Negativity ($\mathcal{L} \mathcal{N}$) is introduced as a convenient and easily interpretable measure of entanglement. For a bipartite state ($\rho$), the ($\mathcal{L} \mathcal{N}$) is defined as: \cite{plenio2005logarithmic}
\begin{equation}
    \label{LN1}
    \mathcal{L} \mathcal{N}(\rho)=\log_2 \Vert\rho^{T_B}\Vert_1 ,
\end{equation}
where $\rho^{T_B}$ is the partial transposition of $\rho$ relative to the subsystem $B$. The trace norm of an operator $\zeta$ is $\Vert\zeta\Vert_1=tr\vert\zeta\vert=tr\sqrt{\zeta}=tr\sqrt{\zeta^{\dagger}\zeta}$. Using the trace norm, we define the negativity of a state $\rho$ as \cite{plenio2005logarithmic}
\begin{equation}
    \label{neg}
    \mathcal{N}(\rho)=\frac{\Vert\rho^{T_B}\Vert_1 -1}{2},
\end{equation}
An equivalent definition of the negativity is written in terms of the eigenvalues $\nu_i$ of $\rho^{T_B}$
\begin{equation}
    \label{neg2}
    \mathcal{N}(\rho)=\sum_i \frac{\vert \nu_i \vert - \nu_i}{2},
\end{equation}
This implies that the logarithmic negativity can be expressed as 
\begin{equation}
    \label{LN2}
    \mathcal{L} \mathcal{N}(\rho)=\log_2\big(2\mathcal{N}(\rho)+1\big).
\end{equation}
\subsection{Relative entropy of coherence}
The coherence of a nonclassical state originates from quantum superpositions, which manifest through the off-diagonal elements of the density matrix in a given basis. For a bipartite quantum state \((\rho)\), coherence can be quantified via the relative entropy of coherence \(\mathcal{C}_r(\rho)\), which serves as a reliable measure satisfying the required coherence criteria. For a bipartite quantum system, \(\mathcal{C}_r(\rho)\) is defined, as reported in \cite{plenio2014quantifying}, by
\begin{equation}\label{coh01}
	\mathcal{C}_r (\rho) = \min_{\xi \in \mathfrak{I}} S(\rho \| \xi),
\end{equation}
where \(\xi\) denotes an incoherent state, \(\mathfrak{I}\) is the set of all incoherent states, and \(\mathcal{S}(\rho)\) represents the von Neumann entropy, given by
\[
\mathcal{S}(\rho) = - \mathrm{tr}(\rho \log \rho)
= \sum_{i=1}^{4} \omega_i \log_2 (\omega_i),
\]
with \(\omega_i\) being the eigenvalues of \(\rho\). The quantity \(\mathcal{C}_r(\rho)\) is a distance-based coherence measure that evaluates coherence through the minimum distance between the quantum state and the set of incoherent states. By exploiting the properties of relative entropy, \(\mathcal{C}_r(\rho)\) can be expressed in the simplified form
\begin{equation}\label{coh02}
	\mathcal{C}_r (\rho) = \mathcal{S}(\rho_{diag}) - \mathcal{S}(\rho),
\end{equation}
where \(\rho_{diag}\) corresponds to the incoherent state obtained by removing all off-diagonal elements of \(\rho\) in the reference basis. It is worth noting that the coherence of any incoherent state \(\xi\) is identically zero.

\section{Fractional-Time Quantum Dynamics of an Organic Molecular Dimer}
\label{sec3}

We study the fractional-time quantum evolution of an organic molecular dimer composed of two perylene bisimide (PBI) chromophores covalently linked by a rigid calix[4]arene spacer. Such molecular architectures are particularly appealing due to their exceptional photostability and the possibility of addressing individual molecules using single-molecule spectroscopy techniques \cite{Hildner2011,Otten2014,Vogelsang2018}. Beyond their optical robustness, PBI-based systems have emerged as promising platforms for molecular-scale quantum technologies, enabling coherent quantum control, the implementation of elementary quantum logic operations, and the manipulation of dissipative entanglement pathways \cite{Briggs2011,Huelga2013}. Remarkably, ultrafast femtosecond control of single-molecule qubits has been demonstrated even at room temperature, underscoring their technological relevance \cite{Kolesov2012}.
\begin{figure}[H]
	\centering
	\includegraphics[width= 0.5\textwidth]{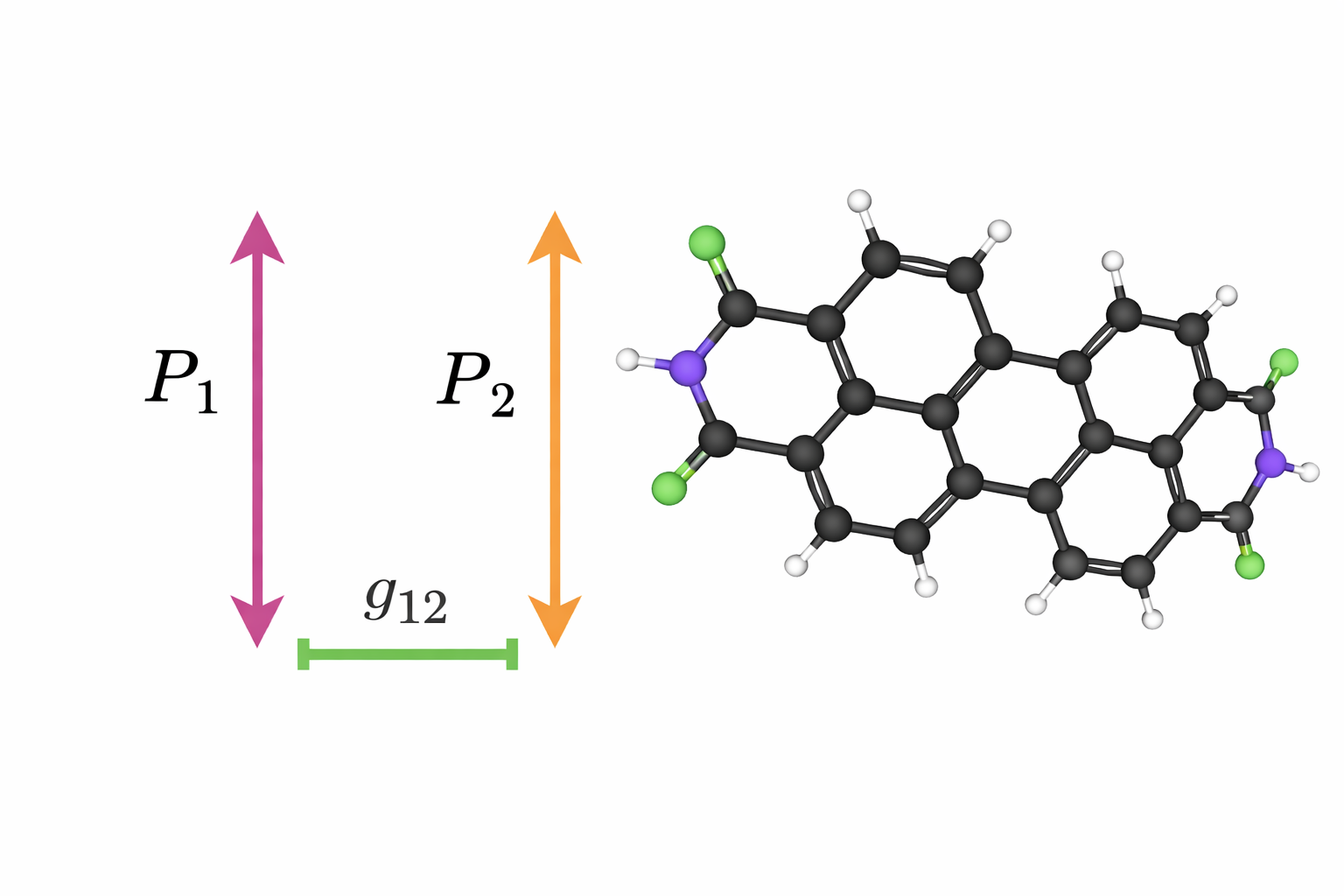}
	\caption{Schematic representation of the molecular dimer system. Two organic molecules are modeled as effective two-level quantum emitters with transition dipole moments \(P_1\) and \(P_2\). The molecules are separated by a distance characterized by the dipole--dipole coupling strength \(g_{12}\), which mediates coherent excitation exchange between them. The chemical structure of the covalently linked molecular dimer is shown on the right.}
	\label{schema_dimer}
\end{figure}
The system consists of two quantum emitters interacting through long-range dipole–dipole coupling. The emitters are separated by a displacement vector $\mathbf{r}_{12}$ and are characterized by transition dipole moments
\[
{\mu}_i=\langle g_i|\mathbf{D}_i|e_i\rangle,
\]
where $\mathbf{D}_i$ denotes the dipole operator of the $i$th molecule, and by individual spontaneous emission rates $\gamma_i$. The collective radiative decay rate $\gamma_{12}$ and the coherent interaction strength $J_{12}$ are given by \cite{Lehmberg1970,Ficek2005}
\begin{align}
\gamma_{12} &= \frac{3}{2}\sqrt{\gamma_1\gamma_2}
\Bigg[
\Big( \hat{\bm{\mu}}_1\!\cdot\!\hat{\bm{\mu}}_2
- (\hat{\bm{\mu}}_1\!\cdot\!\hat{\mathbf{r}}_{12})
(\hat{\bm{\mu}}_2\!\cdot\!\hat{\mathbf{r}}_{12}) \Big)
\frac{\sin \zeta}{\zeta}
\nonumber\\
&\quad+
\Big( \hat{\bm{\mu}}_1\!\cdot\!\hat{\bm{\mu}}_2
-3(\hat{\bm{\mu}}_1\!\cdot\!\hat{\mathbf{r}}_{12})
(\hat{\bm{\mu}}_2\!\cdot\!\hat{\mathbf{r}}_{12}) \Big)
\left(
\frac{\cos \zeta}{\zeta^2}-\frac{\sin \zeta}{\zeta^3}
\right)
\Bigg],\\
J_{12} &= \frac{3}{4}\sqrt{\gamma_1\gamma_2}
\Bigg[
\Big( (\hat{\bm{\mu}}_1\!\cdot\!\hat{\mathbf{r}}_{12})
(\hat{\bm{\mu}}_2\!\cdot\!\hat{\mathbf{r}}_{12})
-\hat{\bm{\mu}}_1\!\cdot\!\hat{\bm{\mu}}_2 \Big)
\frac{\cos \zeta}{\zeta}
\nonumber\\
&\quad+
\Big( \hat{\bm{\mu}}_1\!\cdot\!\hat{\bm{\mu}}_2
-3(\hat{\bm{\mu}}_1\!\cdot\!\hat{\mathbf{r}}_{12})
(\hat{\bm{\mu}}_2\!\cdot\!\hat{\mathbf{r}}_{12}) \Big)
\left(
\frac{\cos \zeta}{\zeta^3}
+\frac{\sin \zeta}{\zeta^2}
\right)
\Bigg],
\end{align}
where $\zeta=nkr_{12}$, $n$ is the refractive index of the surrounding medium, and $k=\nu/c$ is the wave number associated with the molecular transition frequency $\nu$. In a dielectric environment, the individual spontaneous emission rate takes the form
\begin{equation}
\nu_i = n\,\frac{\nu_i^3 |{\mu}_i|^2}{3\pi\varepsilon_0\hbar c^3},
\end{equation}
highlighting its cubic dependence on the transition frequency and the enhancement induced by the optical density of states.

Assuming identical emitters with parallel dipole moments oriented perpendicular to $\mathbf{r}_{12}$ and neglecting external driving fields, the dynamics can be described within the Born–Markov approximation by an effective Hamiltonian \cite{breuer2002theory}
\begin{equation}
H_{\mathrm{eff}} = -\frac{\hbar}{2}\left(\nu_1 \sigma_z^{(1)} + \nu_2 \sigma_z^{(2)}\right)
+ \frac{\hbar V_{12}}{2}
\left(
\sigma_x^{(1)}\sigma_x^{(2)} +
\sigma_y^{(1)}\sigma_y^{(2)}
\right).
\end{equation}
\begin{equation}
H_{\text{dimer}} =
\begin{pmatrix}
-\nu_{0} & 0 & 0 & 0 \\
0 & -\dfrac{\Delta}{2} & V_{12} & 0 \\
0 & V_{12} & \dfrac{\Delta}{2} & 0 \\
0 & 0 & 0 & \nu_{0}
\end{pmatrix},
\end{equation}

where $\nu_{0} = \dfrac{\nu_{1} + \nu_{2}}{2}$ is the mean transition frequency and 
$\Delta = \nu_{1} - \nu_{2}$ denotes the detuning.
The eigenvalues of the Hamiltonian $H_{\text{dimer}}$ and the corresponding eigenstates are given by
\begin{align}
\label{state_evolved}
\epsilon_{1} &= -\nu_{0}, 
& \lvert \Phi_{1} \rangle &= \lvert 00 \rangle, \\
\epsilon_{2} &= -\dfrac{\alpha}{2}, 
& \lvert \Phi_{2} \rangle &= \alpha
\left(
\beta  \lvert 01 \rangle + \lvert 10 \rangle
\right), \\
\epsilon_{3} &= \dfrac{\alpha}{2}, 
& \lvert \Phi_{3} \rangle &= 
\gamma\left(
 \delta \lvert 01 \rangle + \lvert 10 \rangle
\right), \\
\epsilon_{4} &= \nu_{0}, 
& \lvert \Phi_{4} \rangle &= \lvert 11 \rangle.
\end{align}

Here, $\alpha =
\frac{1}{\sqrt{
1+\frac{1}{4}
\left|
\frac{\nu_{1}-\nu_{2}+\sqrt{4V_{12}^{2}+\nu_{1}^{2}-2\nu_{1}\nu_{2}+\nu_{2}^{2}}}
{V_{12}}
\right|^{2}
}}$, $\beta =
-\frac{\nu_{1}-\nu_{2}+\sqrt{4V_{12}^{2}+\nu_{1}^{2}-2\nu_{1}\nu_{2}+\nu_{2}^{2}}}
{2V_{12}}
$,\\ $\gamma =
\frac{1}{\sqrt{
1+\frac{1}{4}
\left|
\frac{\nu_{1}-\nu_{2}-\sqrt{4V_{12}^{2}+\nu_{1}^{2}-2\nu_{1}\nu_{2}+\nu_{2}^{2}}}
{V_{12}}
\right|^{2}
}}$ and $\delta =
\frac{1}{\sqrt{
1+\frac{1}{4}
\left|
\frac{\nu_{1}-\nu_{2}-\sqrt{4V_{12}^{2}+\nu_{1}^{2}-2\nu_{1}\nu_{2}+\nu_{2}^{2}}}
{V_{12}}
\right|^{2}
}}$.

In the resonant case $\nu_{1} = \nu_{2}$, the central eigenstates reduce to symmetric and antisymmetric Bell states.
We assume that the dimer is initially prepared in the Gibbs state of the full Hamiltonian $H_{\text{dimer}}$, corresponding to global thermal equilibrium.

In the computational basis $\{|00\rangle,|01\rangle,|10\rangle,|11\rangle\}$, this Hamiltonian exhibits separable ground and excited states together with coherent superpositions in the single-excitation subspace. In the resonant limit $\nu_1=\nu_2=0$, the interaction term alone governs the dynamics and generates maximally entangled Bell states $|\Psi_{\pm}\rangle=(|10\rangle\pm|01\rangle)/\sqrt{2}$.

To capture memory effects and anomalous temporal behavior beyond standard Markovian dynamics, we describe the system evolution using the time-fractional Schrödinger equation \cite{Naber2004}
\begin{equation}
i^{\tau}\hbar_{\tau}\,
\prescript{C}{}{D}_{t}^{\tau}
\ket{\Psi(t)}
=
H_{\mathrm{eff}}\,\ket{\Psi(t)},
\end{equation}
where $0<\tau\leq 1$ and $\prescript{C}{}{D}_{t}^{\tau}$ denotes the Caputo fractional derivative.  
The formal solution of this equation involves the Mittag--Leffler function,
\begin{equation}
E_{\tau}(x)=\sum_{k=0}^{\infty}\frac{x^k}{\Gamma(\tau k+1)},
\end{equation}
leading to
\begin{equation}
|\Psi(t)\rangle = \sum_{n} C_n
E_{\tau}\!\left(\lambda_n t^{\tau}\right)
|\phi_n\rangle,
\end{equation}
where $\lambda_n$ and $|\phi_n\rangle$ are the eigenvalues and eigenvectors of the Hamiltonian.

The system is initially prepared in a partially coherent superposition
\begin{equation}
|\Psi(0)\rangle = p\,|00\rangle + \sqrt{1-p^2}\,|11\rangle,
\end{equation}

where \(0 \leq p \leq 1\) controls the degree of purity and population imbalance of the state, which in turn affects the quantum correlations and coherence. 

This initial state can also be decomposed in the eigenbasis of the Hamiltonian:
\begin{equation}
\label{init_decomposed}
|\Psi(0)\rangle = \sum_{j=1}^{4} \mathcal{C}_j |\phi_j\rangle,
\end{equation}
from which the coefficients \(\mathcal{C}_j\) can be extracted. Explicitly, they are given by:
\begin{equation}
\label{coeffs}
\begin{aligned}
\mathcal{C}_1 &= 0, \quad \mathcal{C}_2 = \frac{
p - \delta \sqrt{1 - p^{2}}
}{
\gamma\, \beta
- \alpha\, \delta
},   \quad \mathcal{C}_3 =\frac{
1}{\gamma}(
\sqrt{1 - p^{2}}-{\gamma} *\mathcal{C}_2),\quad \mathcal{C}_4 &= 0.
\end{aligned}
\end{equation}

where \(\alpha\), \(\beta\), \(\gamma\), and \(\delta\) are constants related to the eigenstructure of \({H}_{\text{dimer}}\).Substituting these coefficients into Eq.(\ref{state_evolved}) gives:
\begin{equation}
\label{Psi_final_reform}
\begin{aligned}
|\Psi(t, \tau)\rangle &= \chi_1(t,\tau)\, |00\rangle 
+ \chi_2(t,\tau)\, |01\rangle 
+ \chi_3(t,\tau)\, |10\rangle + \chi_4(t,\tau)\, |11\rangle.
\end{aligned}
\end{equation}

with components:
\begin{align}
\chi_1(t,\tau) &= 0, \\
\chi_2(t,\tau) &= \mathcal{C}_2 \, E_\tau(\Phi_2 t^\tau) \cdot \alpha \cdot \beta + \mathcal{C}_3 \, E_\tau(\Phi_3 t^\tau) \cdot \delta \cdot \gamma, \\
\chi_3(t,\tau) &=\mathcal{C}_2 \, E_\tau(\Phi_2 t^\tau) \cdot \alpha  + \mathcal{C}_3 \, E_\tau(\Phi_3 t^\tau) \cdot \gamma, \\
\chi_4(t,\tau) &= 0.
\end{align}
The corresponding time-dependent density matrix.
\begin{equation}
\rho(t,\tau)=\frac{1}{\mathcal{N}}
\begin{pmatrix}
|\chi_1|^2 & 0 & 0 & \chi_1\chi_4^{*}\\
0 & 0 & 0 & 0\\
0 & 0 & 0 & 0\\
\chi_4\chi_1^{*} & 0 & 0 & |\chi_4|^2
\end{pmatrix},
\end{equation}
where $\mathcal{N}=|\chi_1|^2+|\chi_4|^2$ ensures normalization. This unified framework enables a systematic investigation of how fractional-time dynamics influences quantum coherence,  and entanglement in organic molecular dimers.

\section{Results and Discussion \label{sec4}}  

\begin{figure}[H]
\centering
\begin{subfigure}[b]{0.32\textwidth}
    \includegraphics[width=\textwidth]{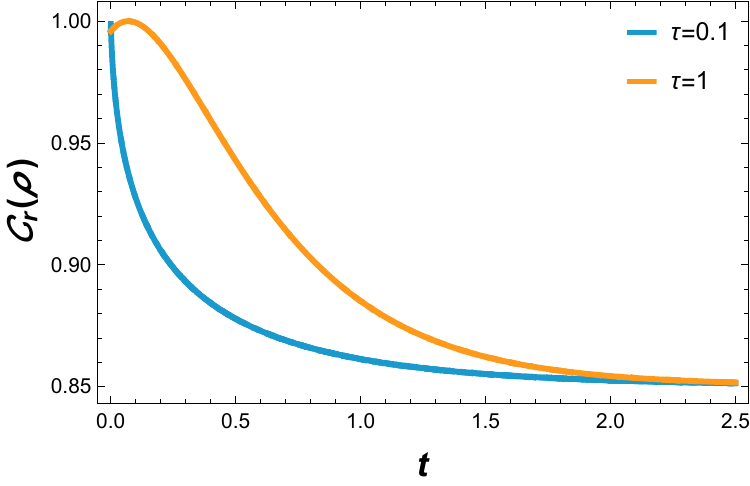}
    \caption{}
    \label{fig1a}
\end{subfigure}
\hfill
\begin{subfigure}[b]{0.32\textwidth}
    \includegraphics[width=\textwidth]{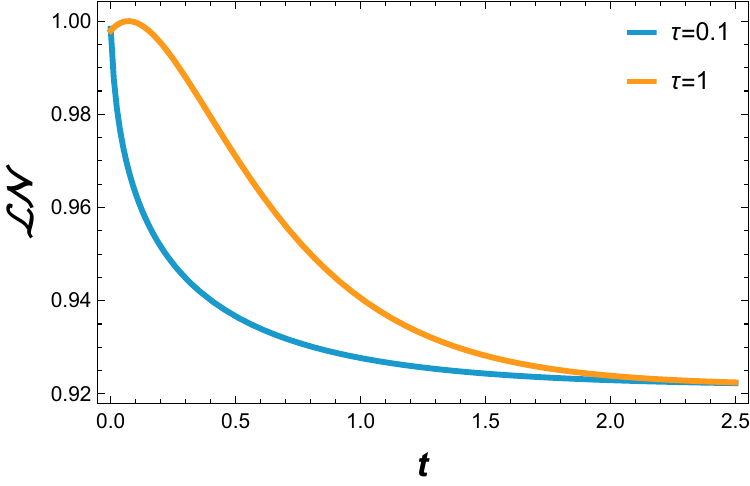}
    \caption{}
    \label{fig1b}
\end{subfigure}
\hfill
\begin{subfigure}[b]{0.32\textwidth}
    \includegraphics[width=\textwidth]{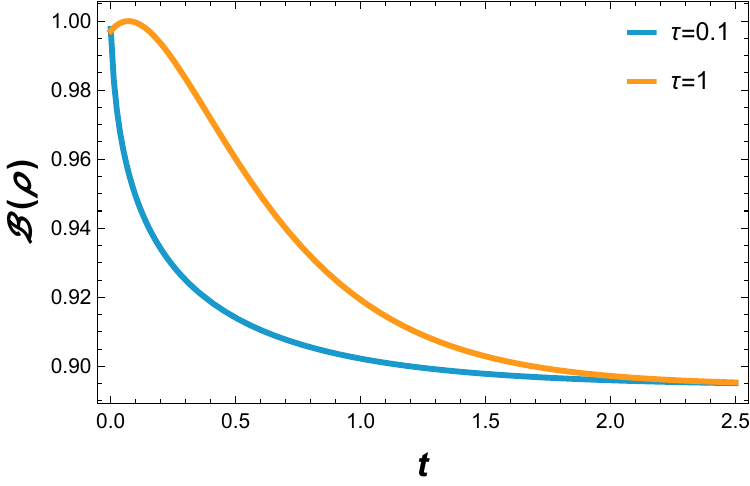}
    \caption{}
    \label{fig1c}
\end{subfigure}
\begin{subfigure}[b]{0.32\textwidth}
    \includegraphics[width=\textwidth]{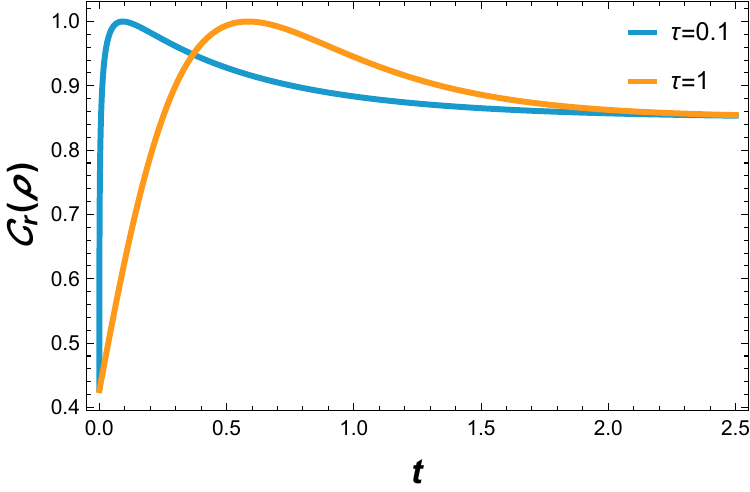}
    \caption{}
    \label{fig1d}
\end{subfigure}
\hfill
\begin{subfigure}[b]{0.32\textwidth}
    \includegraphics[width=\textwidth]{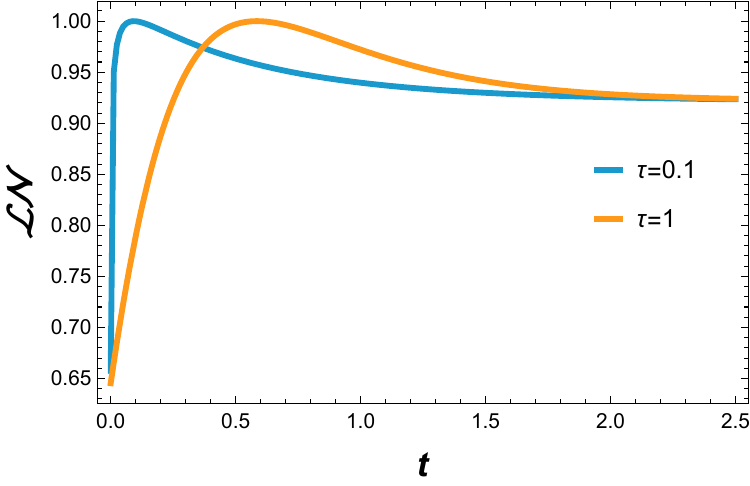}
    \caption{}
    \label{fig1e}
\end{subfigure}
\hfill
\begin{subfigure}[b]{0.32\textwidth}
    \includegraphics[width=\textwidth]{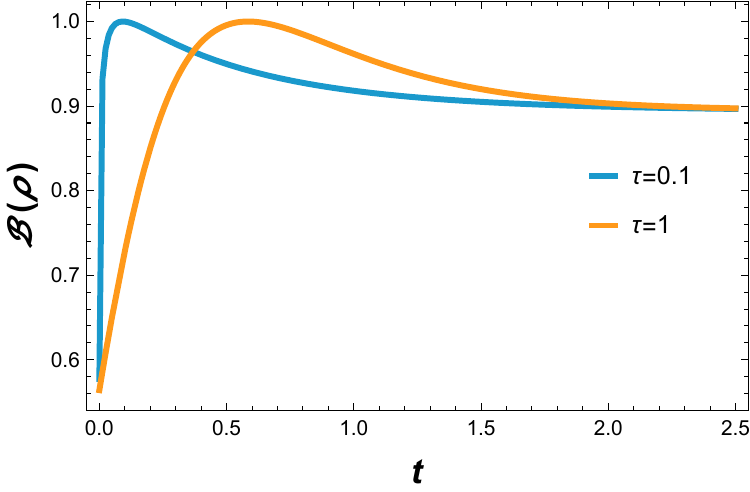}
    \caption{}
    \label{fig1f}
\end{subfigure}
\caption{Comparative plots showing the time evolution of 
$\mathcal{C}_{r}(\rho)$~(\subref{fig1a}--\subref{fig1d}), 
$\mathcal{LN}$~(\subref{fig1b}--\subref{fig1e}), and 
$\mathcal{B}(\rho)$~(\subref{fig1c}--\subref{fig1f}) 
as functions of time $t$, for two values of the fractional parameter: 
$\tau = 0.1$ (fractional dynamics) and $\tau = 1$ (standard Schrödinger evolution).
The results are obtained by fixing 
$\nu_{1} = 1$, $\nu_{2} = 2$, and $V_{12} = 1$, 
for $p = \frac{1}{\sqrt{2}}$ \textit{(top panels)} and 
$p = \frac{1}{\sqrt{6}}$ \textit{(bottom panels)}.}
\label{figure1}
\end{figure}

The obtained dynamics of quantum resources, namely coherence, entanglement and Bell nonlocality, for a PBI molecular dimer undergoing fractional time evolution are plotted in Fig.~(\ref{figure1}). A comparison between the fractional regime ($\tau$ = 0.1) and the standard Schrödinger evolution ($\tau$ = 1) shows that memory effects peculiar to the fractional formalism dramatically reduce the decoherence pace for an initially strongly correlated state ($p = 1/\sqrt{2}$) and boost the dynamical creation of quantum correlations when starting from an initially weakly coherent state ($p = 1/\sqrt{6}$). Moreover, the dependence on the transition frequency $\nu_i$ and the dipole-dipole interaction strength  $V_{12}$ tunes both the preservation and creation of quantum resources, further confirming the role of $\tau$ as a main control parameter of the system's non Markovian features. These results attest the potential of fractional dynamics for robust quantum information control by means of memoryassisted environments able to stabilize and optimize quantum correlations.
In particular, the increased persistence of entanglement and Bell non-locality in the fractional regime establishes the constructive role that memory effects can play in protecting quantum resources against decoherence. These findings therefore support the view that the joint engineering of dynamical parameters and the effective environment may provide a viable pathway toward extended coherence times and improved performance of molecular quantum devices. In this sense, the fractional approach represents a promising framework for devising decoherence-resistant quantum platforms in which genuine non-Markovian effects are actively exploited for the processing and storage of quantum information.

\begin{figure}[H]
\centering
\begin{subfigure}[b]{0.32\textwidth}
    \includegraphics[width=\textwidth]{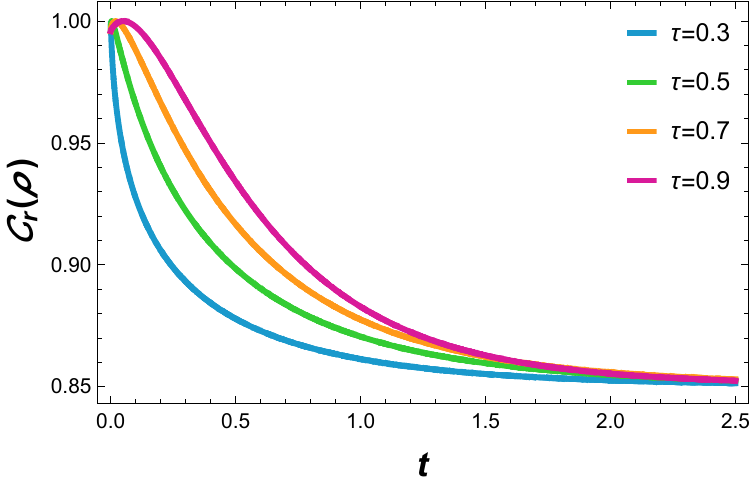}
    \caption{}
    \label{fig2a}
\end{subfigure}
\hfill
\begin{subfigure}[b]{0.32\textwidth}
    \includegraphics[width=\textwidth]{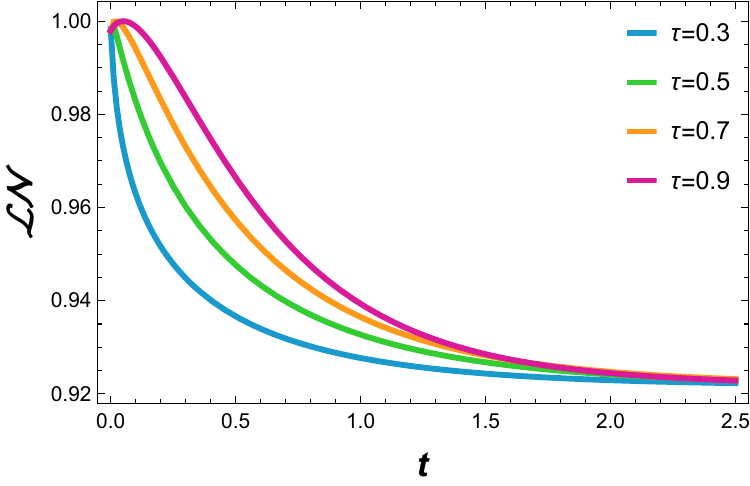}
    \caption{}
    \label{fig2b}
\end{subfigure}
\hfill
\begin{subfigure}[b]{0.32\textwidth}
    \includegraphics[width=\textwidth]{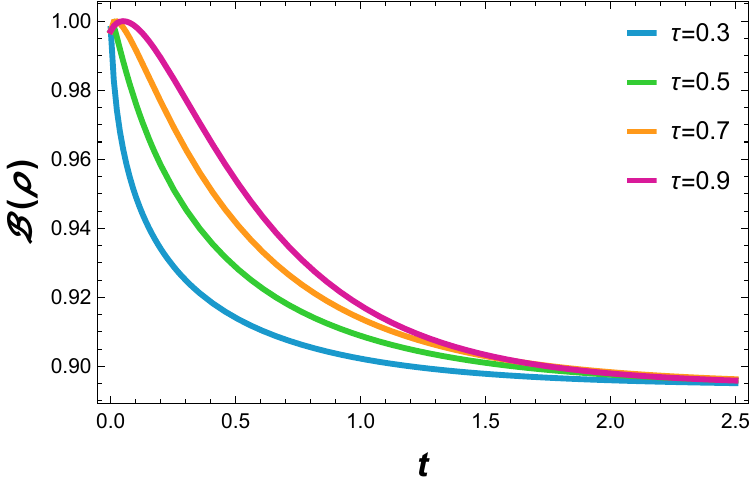}
    \caption{}
    \label{fig2c}
\end{subfigure}
\begin{subfigure}[b]{0.32\textwidth}
    \includegraphics[width=\textwidth]{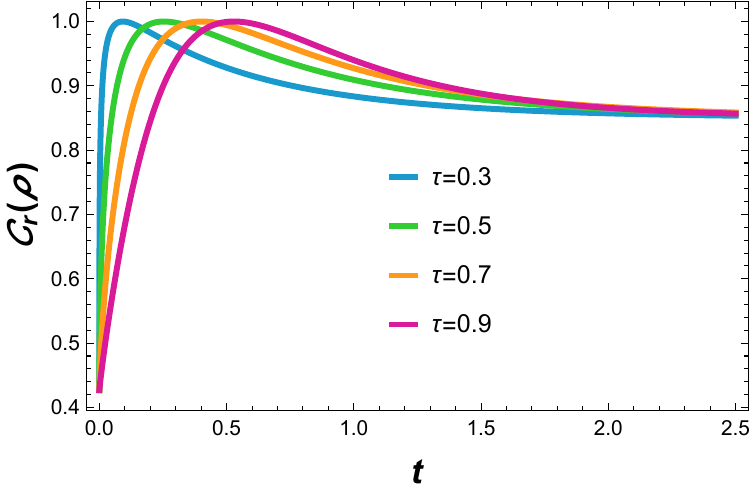}
    \caption{}
    \label{fig2d}
\end{subfigure}
\hfill
\begin{subfigure}[b]{0.32\textwidth}
    \includegraphics[width=\textwidth]{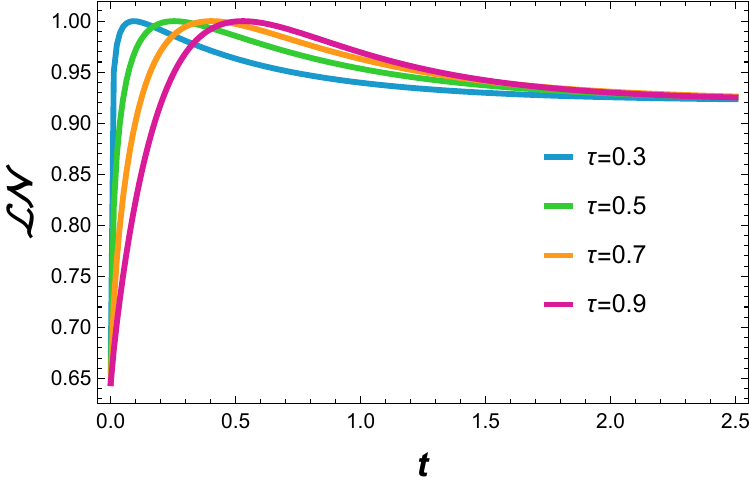}
    \caption{}
    \label{fig2e}
\end{subfigure}
\hfill
\begin{subfigure}[b]{0.32\textwidth}
    \includegraphics[width=\textwidth]{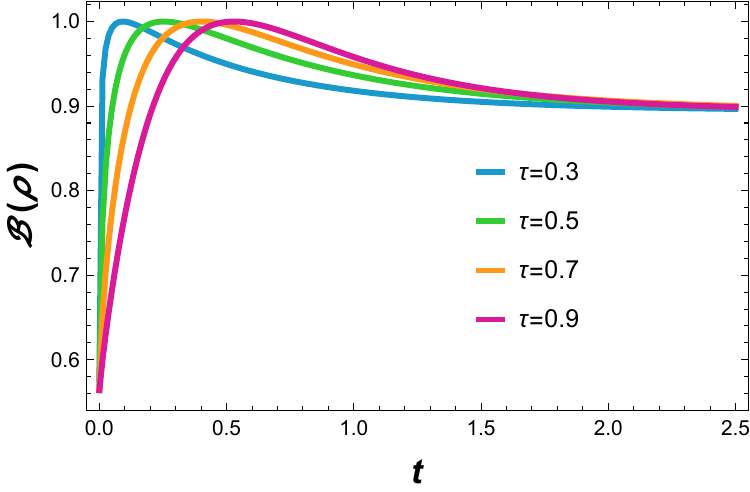}
    \caption{}
    \label{fig2f}
\end{subfigure}
\caption{Plots illustrating $\mathcal{C}_{r}(\rho)$~(\subref{fig2a}--\subref{fig2d}), 
$\mathcal{LN}$~(\subref{fig2b}--\subref{fig2e}), and 
$\mathcal{B}(\rho)$~(\subref{fig2c}--\subref{fig2f}) 
as functions of time $t$ for various values of the fractional parameter \(\tau\),  
$\nu_{1} = 1$, $\nu_{2} = 2$, and $V_{12} = 1$, 
for $p = \frac{1}{\sqrt{2}}$ \textit{(top panels)} and 
$p = \frac{1}{\sqrt{6}}$ \textit{(bottom panels)}.}

\label{figure2}
\end{figure}

Figure~\ref{figure2} illustrates the influence of the fractional parameter \((\tau)\) on the temporal dynamics of three key quantum resources—quantum coherence \(\mathcal{C}_r(\rho)\), entanglement quantified by the logarithmic negativity \(\mathcal{LN}\), and Bell nonlocality \(\mathcal{B}(\rho)\)—for a PBI molecular dimer under fixed dipole–dipole coupling conditions \((\nu_1 = 1,\ \nu_2 = 2,\ V_{12} = 1)\).

When the maximally entangled initial state \((p = 1/\sqrt{2})\) is considered, the upper plots indicate that all resources in the quantum realm are subject to a monotonic decay over time.
However, the rate of decay is strongly governed by the fractional parameter \(\tau\).
In particular, for smaller values of \(\tau\) (e.g., \(\tau = 0.3\)),
the coherence \(\mathcal{C}_r(\rho)\) reaches the reference value
\(\mathcal{C}_r(\rho) = 0.85\) at a later time, \(t = 2\),
whereas for larger values of \(\tau\) (as in the case of \(\tau = 0.9\)),
the same value is reached much earlier, at \(t = 1.5\).
This clearly shows that decreasing the value of \(\tau\) helps slow down the degradation of quantum resources. From a physical perspective, the parameter \(\tau\)
effectively controls the strength of memory effects in the temporal evolution.
Values of \(\tau\) close to unity correspond to a quasi-Markovian regime,
characterized by a rapid loss of system memory and a
predominantly exponential relaxation.
However, for small values of \(\tau\) (\(\tau \ll 1\)),
the system enters a non-Markovian regime,
in which long-time temporal correlations persist.
Information about past system states is partially retained.
As a result, the backflow of information from the environment to the system becomes more evident,
thereby suppressing decoherence and improving the robustness of quantum correlations.
In conclusion, \(\tau\) is found to be a useful control parameter,
enabling an extended preservation of quantum resources and providing a promising path towards memory-based quantum information processing~\cite{Hilfer2000,chhieb2024time}.

For a partially entangled initial state \((p = 1/\sqrt{6})\), it can be seen from the lower panels (\ref{fig2d}–\ref{fig2f}) that a different kind of dynamical evolution takes place. The quantum resources, which grow, reach a maximum, and then diminish, exhibit an increased rate of growth and maximum values for \(\mathcal{C}_r(\rho)\), \(\mathcal{LN}\), and \(\mathcal{B}(\rho)\) with reduced values of the fractional parameter, \(\tau\). The parameter, \(\tau\), serves as a catalyst for the coherent creation of the entanglement from the weakly entangled state, hence contributing to the dynamics of the creation of quantum resources. It has also been shown that the values of \(\mathcal{C}_r(\rho)\), \(\mathcal{LN}\), and \(\mathcal{B}(\rho)\) remain higher for reduced values of the parameter, \(\tau\). The partial preservation of freshly generated coherences allows for greater efficiency in the generation and temporary stabilization of quantum resources before relaxation mechanisms become dominant. These features establish that this fractional approach is capable not only of sustaining previously existing quantum resources in the system, but also of enhancing their dynamical generation in an interacting setting~\cite{chhieb2025fractional,el2022dynamics}. These results establish that the value of the fractional order, denoted as \((\tau)\), is an effective controlling factor for the dynamics of quantum correlations in molecular systems. From a physical perspective, the value of \(\tau\) may be related to the properties of the environment, such as the phononic density of states, the rigidity of the supporting matrix, or the role of disorder, and in the end, all these factors are related to information preservation in the system. For PBI dimers, or in general for quantum systems, the obtained results suggest that, in order to control the dynamics of quantum resources, an indirect way to act on the system could be through the influence of the environment, making non-Markovianity an important tool in the preservation and creation of quantum correlations.

\begin{figure}[H]
\centering
\begin{subfigure}[b]{0.32\textwidth}
    \includegraphics[width=\textwidth]{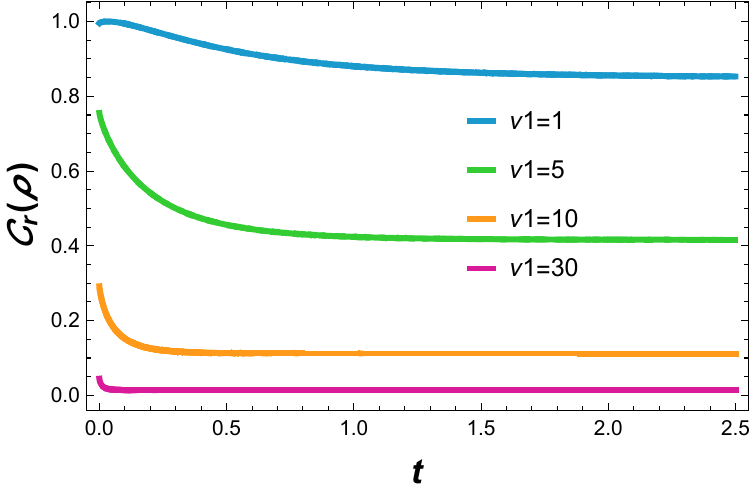}
    \caption{}
    \label{fig3a}
\end{subfigure}
\hfill
\begin{subfigure}[b]{0.32\textwidth}
    \includegraphics[width=\textwidth]{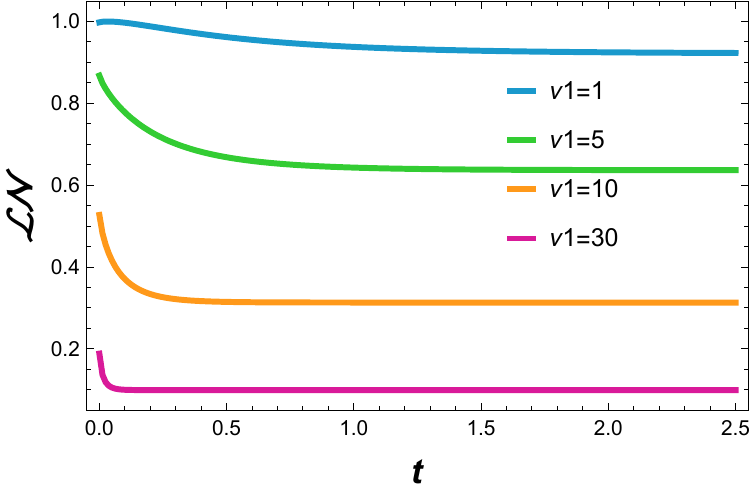}
    \caption{}
    \label{fig3b}
\end{subfigure}
\hfill
\begin{subfigure}[b]{0.32\textwidth}
    \includegraphics[width=\textwidth]{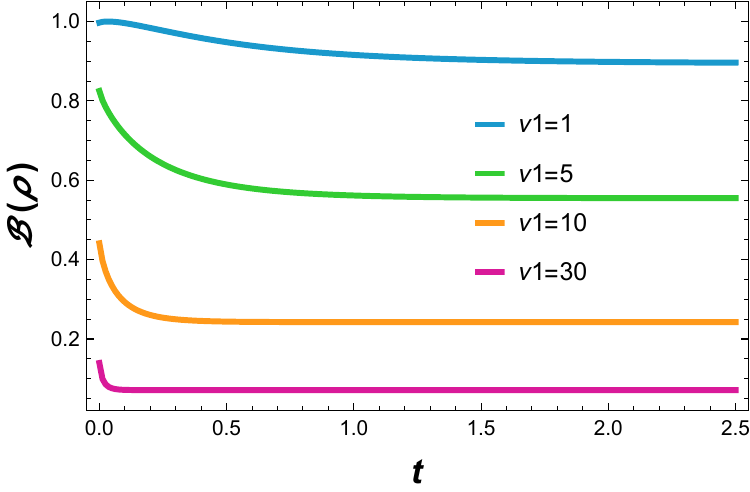}
    \caption{}
    \label{fig3c}
\end{subfigure}
\begin{subfigure}[b]{0.32\textwidth}
    \includegraphics[width=\textwidth]{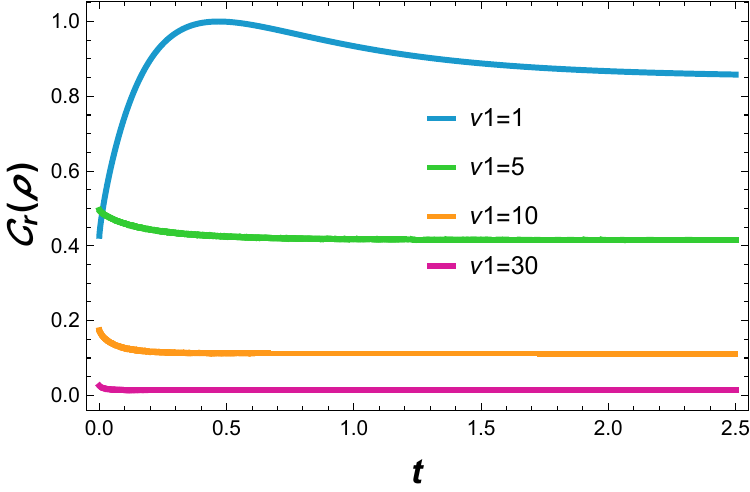}
    \caption{}
    \label{fig3d}
\end{subfigure}
\hfill
\begin{subfigure}[b]{0.32\textwidth}
    \includegraphics[width=\textwidth]{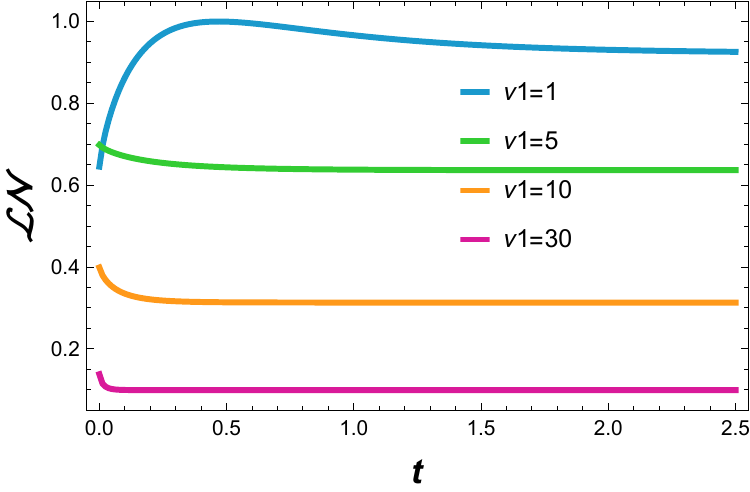}
    \caption{}
    \label{fig3e}
\end{subfigure}
\hfill
\begin{subfigure}[b]{0.32\textwidth}
    \includegraphics[width=\textwidth]{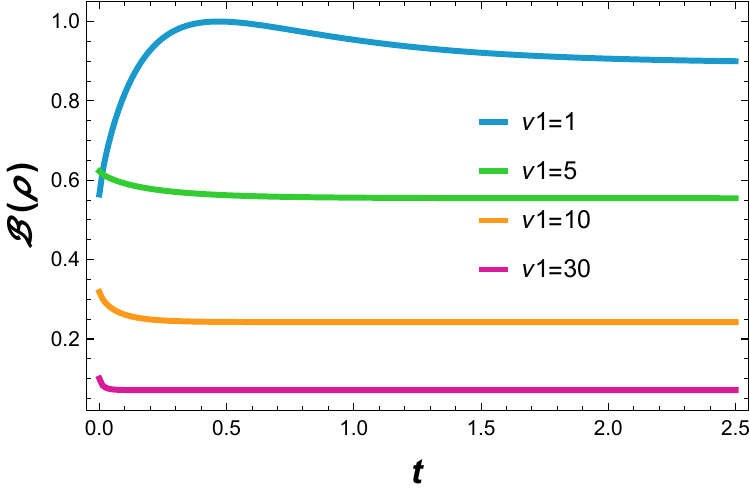}
    \caption{}
    \label{fig3f}
\end{subfigure}
\caption{Plots illustrating Concurrence $\mathcal{C}_{r}(\rho)$~(\subref{fig3a}--\subref{fig3d}), 
$\mathcal{LN}$~(\subref{fig3b}--\subref{fig3e}), and 
$\mathcal{B}(\rho)$~(\subref{fig3c}--\subref{fig3f}) 
transition frequency of site 1 $\nu_{1}$, withe $\tau = 0.8$, $\nu_{2} = 2$, and $V_{12} = 1$, 
for $p = \frac{1}{\sqrt{2}}$ \textit{(top panels)} and 
$p = \frac{1}{\sqrt{6}}$ \textit{(bottom panels)}.}

\label{figure3}
\end{figure} 

Figure~\ref{figure3} illustrates the dependence of three key quantum resources—quantum coherence \(\mathcal{C}(\rho)\), logarithmic negativity \(\mathcal{LN}\), and Bell nonlocality \(\mathcal{B}(\rho)\)—on the transition frequency of site~1, \(\nu_1\), in a time-fractional PBI dimer system with fixed parameters \(\tau = 0.8\), \(\nu_2 = 2\), and \(V_{12} = 1\). The analysis is carried out for two different initial state purities: \(p = 1/\sqrt{2}\) (upper panels, with high initial coherence) and \(p = 1/\sqrt{6}\) (lower panels, with reduced initial coherence).

For the highly coherent initial state (\(p = 1/\sqrt{2}\), upper panels \ref{fig3a}–\ref{fig3c}), all three quantum resources exhibit a gradual decrease as \(\nu_1\) increases. This behavior clearly indicates that frequency detuning between the two molecular sites diminishes quantum correlations. When \(\nu_1\) deviates significantly from \(\nu_2\), the resulting energy mismatch reduces the efficiency of the dipole–dipole interaction \(V_{12}\) in mediating coherent excitation exchange, leading to a reduction in quantum coherence (\(\mathcal{C}(\rho)\) and \(\mathcal{LN}\)) as well as in nonlocality (Bell violation)~\cite{Ficek2005,breuer2002theory}. The faster decay of Bell nonlocality compared to entanglement measures highlights that nonlocal correlations are more sensitive to frequency detuning than bipartite entanglement, in agreement with the established hierarchy of quantum correlations, where nonlocality represents a particularly fragile resource~\cite{Horodecki2009}.

For the less coherent initial state (\(p = 1/\sqrt{6}\), lower panels \ref{fig3d}–\ref{fig3f}), the dynamics exhibit a pronounced non-monotonic behavior, where quantum resources first increase with \(\nu_1\), reach a well-defined optimal maximum, and subsequently decrease. This behavior demonstrates that frequency detuning can act as an efficient dynamical mechanism for enhancing quantum correlation generation when the system is initialized in a partially coherent state. The initial enhancement arises from the fact that a moderate detuning allows a more favorable balance between the coherent interaction \(V_{12}\) and the local energy contributions, thereby optimizing energy-transfer pathways and temporarily amplifying entanglement and nonlocality through fractional dynamics~\cite{el2024entanglement}. The location of the maximum identifies an optimal detuning condition that maximizes the competition between local dynamical evolution and interaction-induced coherence, whereas beyond this point further detuning increasingly suppresses quantum correlations due to growing energy mismatch. These findings carry several important physical implications: first, the transition frequency \(\nu_1\), or equivalently the detuning \(\Delta = \nu_1 - \nu_2\), constitutes a powerful external control parameter that can be experimentally tuned via electric or strain fields to either enhance or suppress quantum correlations depending on the initial state preparation; second, the contrasting trends observed for different values of \(p\) underscore the central role of initial-state engineering, indicating that while near-resonant conditions (\(\nu_1 \approx \nu_2\)) favor highly coherent states, partially coherent states can significantly benefit from a finite detuning; third, for \(\tau = 0.8\), corresponding to an intermediate memory regime, the observed non-monotonic behavior reveals a cooperative interplay between fractional memory effects and frequency detuning, enabling a transient enhancement of quantum correlations that is absent in purely Markovian dynamics (\(\tau = 1\))~\cite{Naber2004}; finally, from a technological perspective, since transition frequencies in molecular and solid-state quantum devices can be chemically engineered or externally modulated, dynamic frequency tuning emerges as a promising strategy for the transient amplification of entanglement and other quantum resources, particularly in platforms initialized in partially coherent states.

\begin{figure}[H]
\centering
\begin{subfigure}[b]{0.32\textwidth}
    \includegraphics[width=\textwidth]{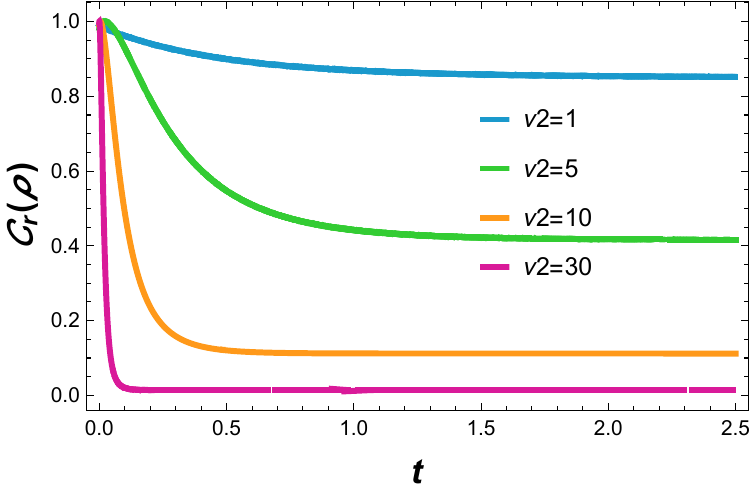}
    \caption{}
    \label{fig4a}
\end{subfigure}
\hfill
\begin{subfigure}[b]{0.32\textwidth}
    \includegraphics[width=\textwidth]{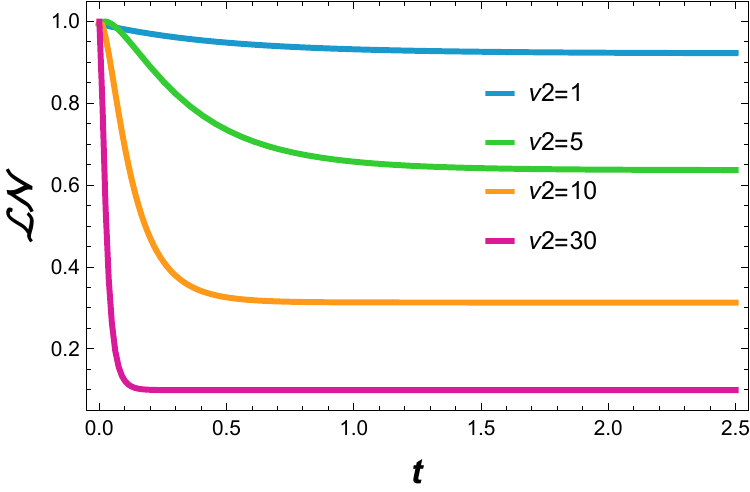}
    \caption{}
    \label{fig4b}
\end{subfigure}
\hfill
\begin{subfigure}[b]{0.32\textwidth}
    \includegraphics[width=\textwidth]{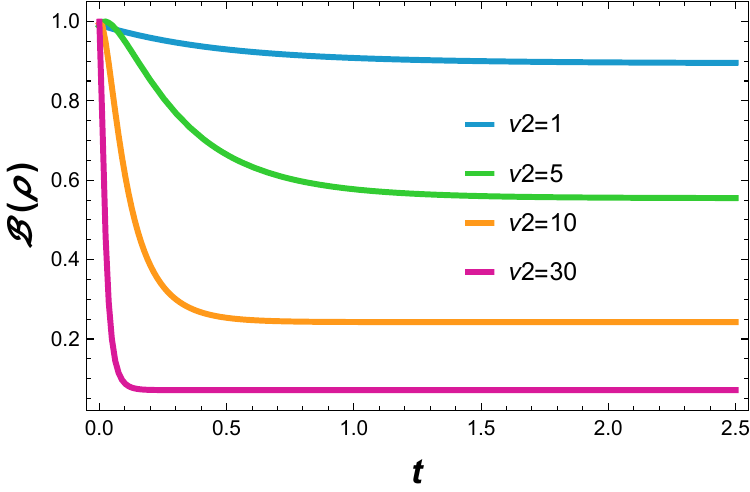}
    \caption{}
    \label{fig4c}
\end{subfigure}
\begin{subfigure}[b]{0.32\textwidth}
    \includegraphics[width=\textwidth]{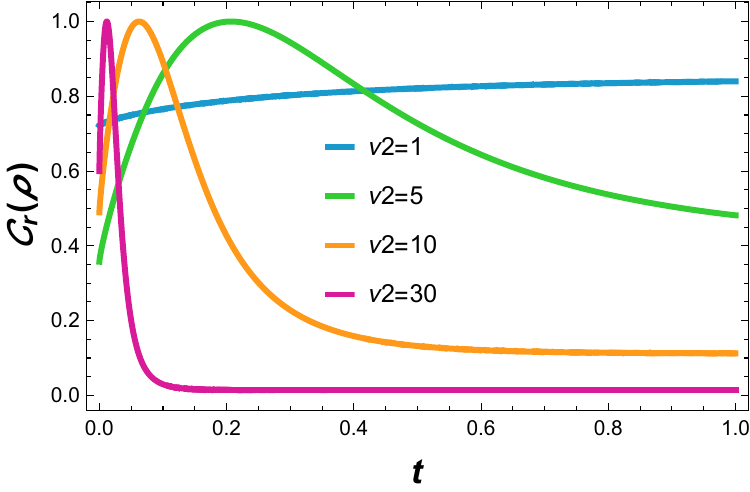}
    \caption{}
    \label{fig4d}
\end{subfigure}
\hfill
\begin{subfigure}[b]{0.32\textwidth}
    \includegraphics[width=\textwidth]{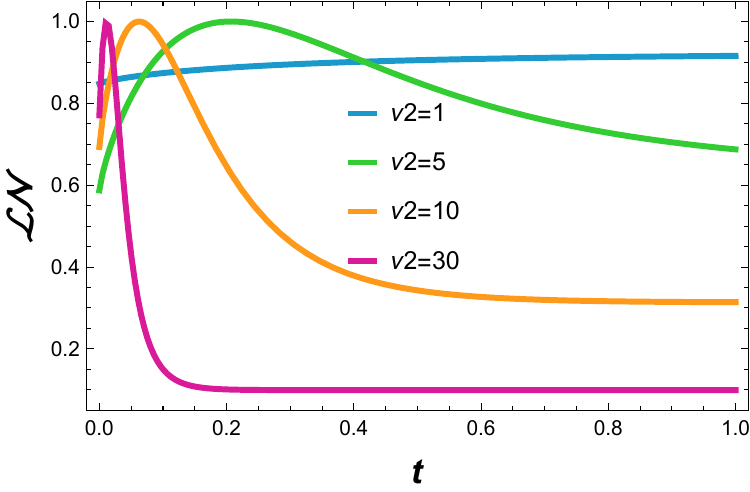}
    \caption{}
    \label{fig4e}
\end{subfigure}
\hfill
\begin{subfigure}[b]{0.32\textwidth}
    \includegraphics[width=\textwidth]{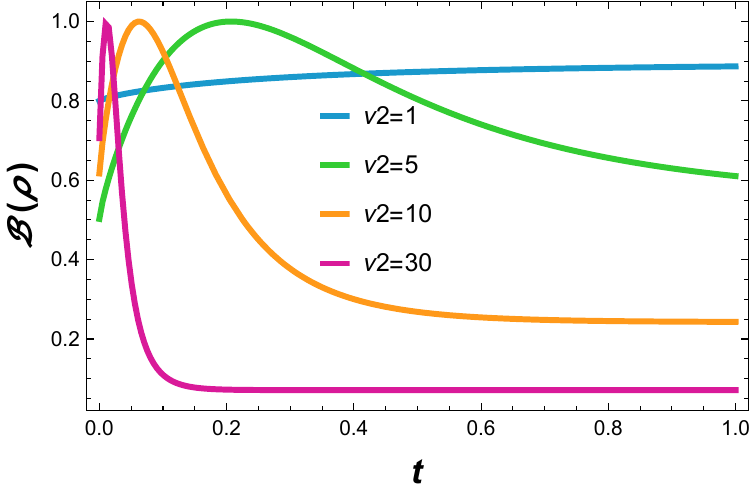}
    \caption{}
    \label{fig4f}
\end{subfigure}
\caption{Plots illustrating $\mathcal{C}_{r}(\rho)$~(\subref{fig4a}--\subref{fig4d}), 
$\mathcal{LN}$~(\subref{fig4b}--\subref{fig4e}), and 
$\mathcal{B}(\rho)$~(\subref{fig4c}--\subref{fig4f}) 
as functions of time $t$ for various values of  the transition frequency of site 2 $\nu_{2}$. withe $\tau= 0.8$, $\nu_{1} = 2$, and $V_{12} = 1$, 
for $p = \frac{1}{\sqrt{2}}$ \textit{(top panels)} and 
$p = \frac{1}{\sqrt{6}}$ \textit{(bottom panels)}.}

\label{figure4}
\end{figure}
Figure~\ref{figure4} reveals the determining influence of the transition frequency of site~2, $\nu_2$, on the dynamics of quantum resources in a fractional-memory regime ($\tau = 0.8$). For a high-purity initial state ($p = 1/\sqrt{2} \approx 0.71$), $\mathcal{C}_r(\rho)$ and entanglement $\mathcal{LN}$ exhibit a monotonic decline, whose rate is directly correlated with the detuning $|\nu_2 - \nu_1|$. For example, under zero detuning ($\nu_2 = 2$), coherence remains above $0.5$ for a duration $t > 5$ time units, whereas a strong detuning ($\nu_2 = 4$) causes it to drop below $0.2$ as early as $t \approx 2$. Bell nonlocality $\mathcal{B}(\rho)$, on the other hand, shows heightened sensitivity, falling below the classical threshold of $2$ more rapidly than the other measures, illustrating its more fragile nature. Conversely, for a partially mixed initial state ($p = 1/\sqrt{6}$), a non-monotonic generation behavior is observed: all Three resources raise from the initially small values to attain the important maximum. The height of the maximum and the time for its attainment, especially with respect to the detuning, have proved to be optimized by the value $
u_2 \approx 3$, for which the entanglement generated is maximized, increasing the value of $\mathcal{LN}$ to nearly $0.4$ from the initial value that is close to $0$. This mechanism is attributed to the coherent interaction $V_{12}$ which, in the presence of system memory ($\tau < 1$), allows for efficient accumulation of correlations before relaxation takes over. Thus, $\nu_2$ acts as a spectral control parameter that, depending on the initial preparation, can either preserve existing resources by maintaining near-resonance ($\nu_2 \approx \nu_1$), or dynamically generate them via an optimal detuning. This dual role, coupled with the stabilizing effect of fractional dynamics, opens concrete prospects for the active control of quantum information in real molecular architectures~\cite{Fassioli2014,Breuer2009}.

\begin{figure}[H]
\centering
\begin{subfigure}[b]{0.32\textwidth}
    \includegraphics[width=\textwidth]{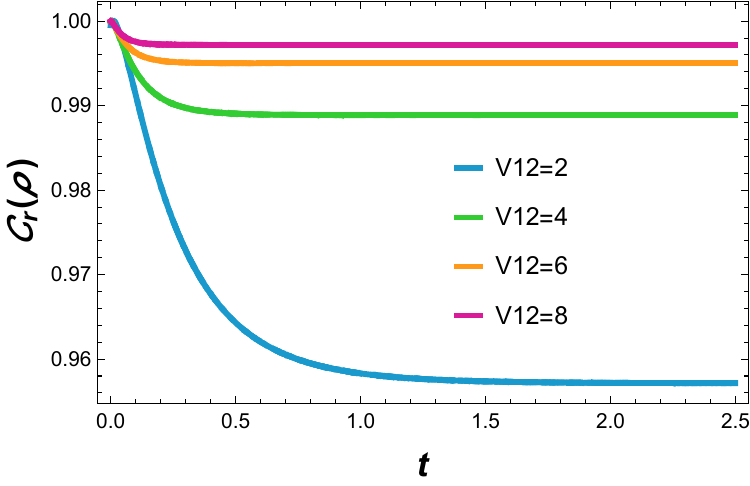}
    \caption{}
    \label{fig5a}
\end{subfigure}
\hfill
\begin{subfigure}[b]{0.32\textwidth}
    \includegraphics[width=\textwidth]{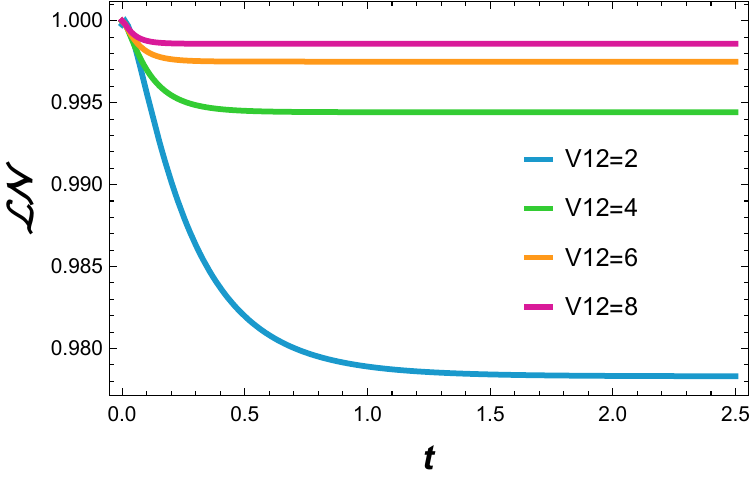}
    \caption{}
    \label{fig5b}
\end{subfigure}
\hfill
\begin{subfigure}[b]{0.32\textwidth}
    \includegraphics[width=\textwidth]{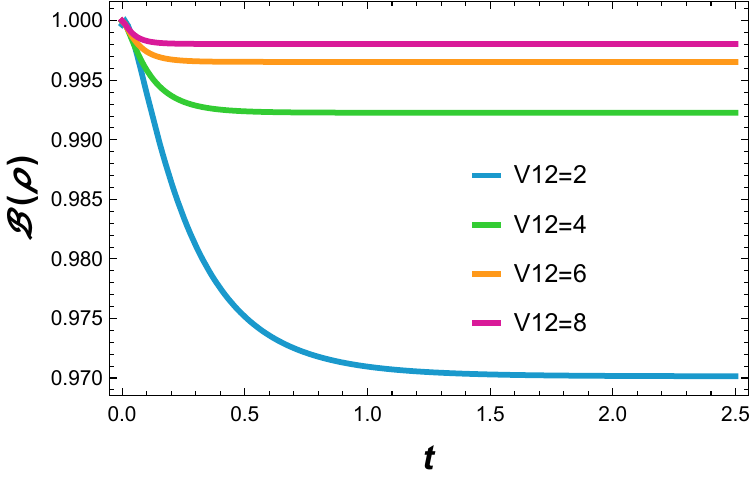}
    \caption{}
    \label{fig5c}
\end{subfigure}
\begin{subfigure}[b]{0.32\textwidth}
    \includegraphics[width=\textwidth]{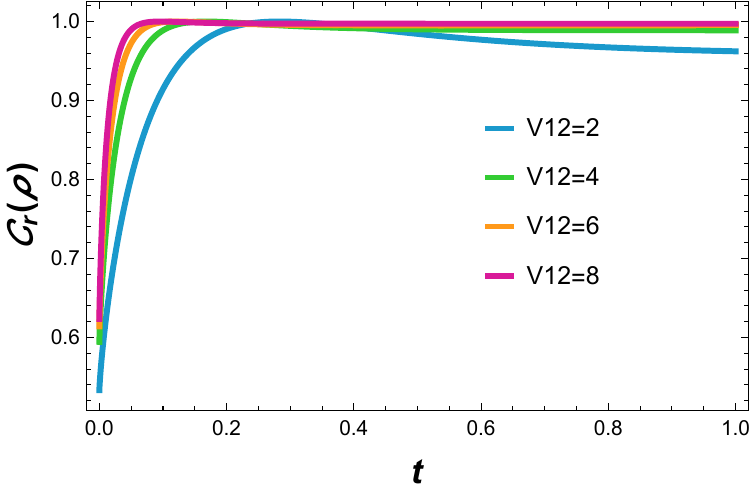}
    \caption{}
    \label{fig5d}
\end{subfigure}
\hfill
\begin{subfigure}[b]{0.32\textwidth}
    \includegraphics[width=\textwidth]{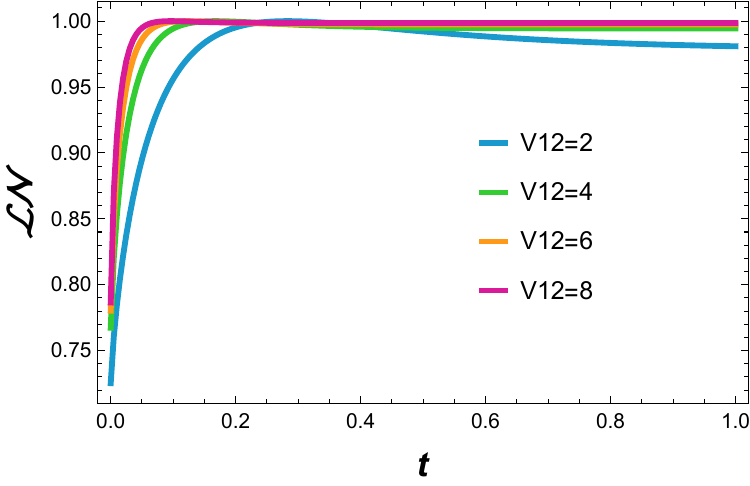}
    \caption{}
    \label{fig5e}
\end{subfigure}
\hfill
\begin{subfigure}[b]{0.32\textwidth}
    \includegraphics[width=\textwidth]{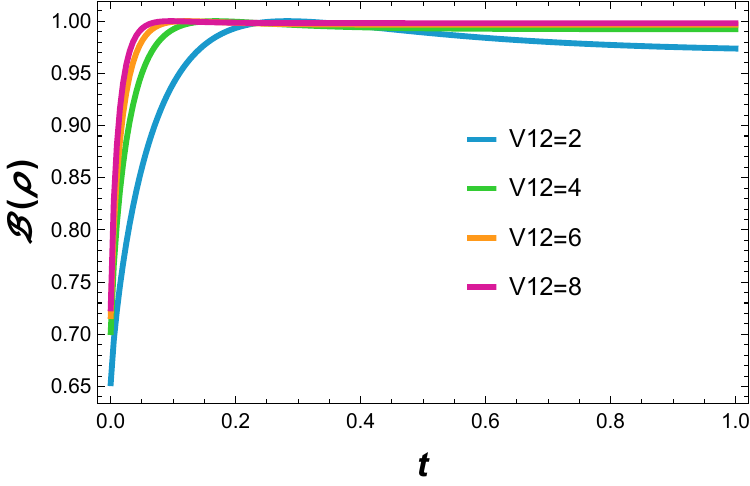}
    \caption{}
    \label{fig5f}
\end{subfigure}
\caption{Plots illustrating $\mathcal{C}_{r}(\rho)$~(\subref{fig5a}--\subref{fig5d}), 
$\mathcal{LN}$~(\subref{fig5b}--\subref{fig5e}), and 
$\mathcal{B}(\rho)$~(\subref{fig5c}--\subref{fig5f}) 
as functions of time $t$ for various values of  the interaction strength $V_{12}$. withe $\nu_{1} = 1$, $\nu_{2} = 2$, and $\tau = 0.8$, 
for $p = \frac{1}{\sqrt{2}}$ \textit{(top panels)} and 
$p = \frac{1}{\sqrt{6}}$ \textit{(bottom panels)}.}

\label{figure5}
\end{figure}
Figure~\ref{figure5} illustrates the time evolution of quantum coherence
\( \mathcal{C}_r(\rho) \), entanglement quantified by the logarithmic
negativity \( \mathcal{LN} \), and Bell nonlocality
\( \mathcal{B}(\rho) \) in a PBI molecular dimer as a function of the
dipole--dipole interaction strength \( V_{12} \).
The system parameters are fixed at
\( \nu_1 = 1 \), \( \nu_2 = 2 \), and \( \tau = 0.8 \).
Two different initial states are considered: a state with intermediate
purity (\( p = 1/\sqrt{2} \), upper panels ( \ref{fig5a}–\ref{fig5c}) and a state characterized
by lower initial coherence (\( p = 1/\sqrt{6} \)), lower panels  (\ref{fig5d}–\ref{fig5f}).

For the maximally entangled initial state (\( p = 1/\sqrt{2} \)),
all three quantum resources exhibit a gradual decay over time.
However, the decay rate is strongly influenced by the value of
\( V_{12} \).
A stronger coupling significantly slows down the loss of coherence and
entanglement; for instance, when \( V_{12} = 8 \),
the coherence \( \mathcal{C}_r(\rho) \) remains above \( 0.99 \) up to
\( t \simeq 0.5 \).
By contrast, for weaker coupling (\( V_{12} = 2 \)),
However, it goes below \( 0.95 \) even at \( t \simeq 1.5 \).
Bell nonlocality also follows the same trend, but goes beyond the classical
threshold \( \mathcal{B} = 2 \) in the weak coupling regime,
emphasizing the increased sensitivity to the interaction strength and the importance of \( V_{12} \) in managing the time dynamics of quantum correlations. For smaller coherence in the initial state, \( p = \frac{1}{\sqrt{6}} \),
In other words, when the value of $\lambda$ approaches infinity, the system
All three quantum resources initially increase, attain a definite maximimum point, and
then decay more slowly.
This increase becomes more evident as \( V_{12} \) increases, reflecting the relative importance of dipole--dipole interactions to the correlation process.
orption), we see that for \( V_{12} = 8 \), the maximal value of the entanglement attains about \( 0.5 \), which, for \( V_{12} = 2 \), is hardly above \( 0.2 \).
The time taken for the system to attain this peak reduces with the increment in the coupling strength, and this reveals that the role of \( V_{12} \) in the system helps in developing the quantum correlation.
This comes about as a result of the mediating action of \( V_{12} \) in the coherent transfer of excitation between the two sites within the molecule, and this action is further aided by the effects associated with the fractional dynamics (\( \tau = 0.8 \)). The strength of the interaction, \( V_{12} \), therefore emerges as an important controlling factor in the management of both the conservation and the creation of quantum resources. Higher strength not only ensures the conservation of the coherence and entanglement but also improves the creation of correlations within the weak coherent states.
These results highlight the importance of molecular engineering and chromophore spatial architecture in optimizing dipole-dipole interactions and open avenues toward controllable molecular quantum systems that could be useful in quantum computing.

\section{Conclusion\label{sec5}}
In this work, we have explored the fractional-time evolution of quantum coherence, entanglement, and Bell nonlocality in a dimeric PBI molecular system with dipole-dipole coupling. By solving the time-fractional Schrödinger equation, we demonstrated that the fractional order $\tau$ plays a crucial role in governing the temporal behavior of quantum resources. Our results illustrate that for $\tau < 1$, there are strong memory effects in the system, with a slower decoherence process in comparison to standard quantum dynamics for $\tau = 1$. Thus, it is clear that fractional dynamics provides a more general, real-world framework for treating coherence preservation in complex open systems with long-time memory environments. The initial state purity parameter $p$ was seen to play a strong role in producing and preserving entanglement as well as the violation of Bell inequalities. Additionally, the strength of dipole-dipole interaction $V_{12}$ and the respective transition frequencies $\nu_i$ may be carefully controlled to either enhance or suppress quantum entanglement over time, thereby effectively controlling quantum dynamics. 
These findings highlight the perspectives that the fractional approach gives for the design of robust quantum molecular systems, even at ambient and noisy conditions. The present research combines fractional calculus with quantum-information theory in a way that opens a new avenue of investigation concerning the development of strategies managing and controlling quantum states in emerging quantum technologies. Further studies may be developed regarding the extension of the present research to multipartite systems, considering more general non-Markovian effects and exploring different boundary conditions and molecular geometries. In summary, the fractional approach acts as an important and promising method to model and understand quantum dynamics. These insights open the door to further experimental and theoretical investigations aimed at exploiting fractional dynamics for enhanced control of quantum resources in realistic molecular and hybrid quantum systems.

\section*{Appendix A}
Fractional calculus has emerged as a powerful mathematical tool in quantum physics, extending conventional differential operators to non-integer orders. The Riemann-Liouville integral serves as a foundational element in this framework, naturally leading to fractional evolution equations like the fractional Schrödinger equation (Eq.~\ref{FTSE}).The Riemann-Liouville integral of order $\tau > 0$ for a function $F(t)$ is defined as
\begin{equation}
\mathcal{J}^{\tau} F(t) = \frac{1}{\Gamma(\tau)} \int_{0}^{t} (t - s)^{\tau - 1} F(s)\, ds, \qquad t > 0,
\label{eq:RL_integral_new}
\end{equation}
where $\Gamma(\tau)$ is the Gamma function. This integral operator satisfies the composition properties
\begin{equation}
\mathcal{J}^{\tau}\mathcal{J}^{\beta} F(t) = \mathcal{J}^{\tau+\beta} F(t),
\mathcal{J}^{\tau} t^{m} = \frac{\Gamma(m+1)}{\Gamma(\tau+m+1)} t^{\tau+m}.
\end{equation}

Within the Caputo framework, this integral subsequently defines the fractional derivative of order $\\tau$: for a function $F(t)$, it is given by
\begin{equation}
\mathcal{D}^{\tau} F(t) = \mathcal{J}^{n-\tau} \left( \frac{d^{n}}{dt^{n}} F(t) \right)
= \frac{1}{\Gamma(n-\tau)} \int_{0}^{t} \frac{F^{(n)}(s)}{(t-s)^{\tau-n+1}} \, ds,
\end{equation} where $n = \lceil \tau \rceil$. This formula explicitly relates fractional derivatives to integer-order derivatives. A general linear fractional system of order $\tau >0$ then obeys
\begin{equation}
\label{eq:frac_system}
\frac{d^{\tau}}{dt^{\tau}} \mathbf{Y}(t) = \mathbf{A} \mathbf{Y}(t), \qquad \mathbf{Y}(0) = \mathbf{Y}_{0},
\end{equation}
where $\mathbf{Y} \in \mathbb{R}^{n}$, $\mathbf{A} \in \mathbb{R}^{n \times n}$, and the derivatives are taken in the Caputo sense.
The solution of Eq.~(\ref{eq:frac_system}) involves the one-parameter Mittag-Leffler function
\begin{equation}
E_{\tau}(z) = \sum_{k=0}^{\infty} \frac{z^{k}}{\Gamma(\tau k + 1)} .
\end{equation}
A particular solution takes the form
\begin{equation}
\mathbf{Y}(t) = \mathbf{v} \, E_{\tau}(\lambda t^{\tau}),
\end{equation}
where $\mathbf{v}$ is a constant vector and $\lambda$ is related to the eigenvalues of $\mathbf{A}$. Substitution into Eq.~(\ref{eq:frac_system}) yields the eigenvalue equation
\begin{equation}
\mathbf{A} \mathbf{v} = \lambda \mathbf{v} \quad \Longleftrightarrow \quad (\mathbf{A} - \lambda \mathbf{I}_{n}) \mathbf{v} = 0,
\end{equation}
demonstrating that $\lambda$ and $\mathbf{v}$ are spectral components of $\mathbf{A}$. Consequently, the general solution is given by
\begin{equation}
\mathbf{Y}(t) = \sum_{j=1}^{n} c_{j} \, E_{\tau}(\lambda_{j} t^{\tau}) \mathbf{v}_{j},
\end{equation}
with coefficients $c_{j}$ determined by the initial condition $\mathbf{Y}(0)$.

\section*{Declarations}

\subsection*{Funding and/or Conflicts of Interest}
This research received no external funding. The authors declare that they have no conflicts of interest.
\newpage

\printbibliography

\end{document}